\documentclass[pra,preprint,showpacs,floatfix]{revtex4}
\usepackage{graphicx}
\usepackage{psfrag}
\usepackage{amssymb,amsmath}

\begin{document}

\title{Zitterbewegung and the Charge of an Electron}

\author{Basil~S.~Davis}

\email{bdavis22@xula.edu}

\affiliation{Department of Physics and Computer Science, Xavier University of Louisiana, New Orleans, Louisiana 70125, USA}

\begin{abstract}

Dirac's Relativistic Wave Equation implies a measured electron velocity of  $\pm c$  in any direction, in contradiction to Special Relativity and observation. It is shown in this article that this anomalous electron velocity reveals an internal structure of the electron whereby the mass and the charge of the electron cannot be described by the same position operator. The measured velocity of electron mass is always less than $c$ in any direction but charge can be displaced at the speed of light. This speed is realizable only when the electron is in a state that is a superposition of positive and negative energy states, also known as a zitterbewegung state. It is shown that in zitterbewegung it is the charge and not the mass that undergoes rapid spatial oscillation, and that there are measurable consequences of this charge zitterbewegung. Zitterbewegung of charge also occurs in an entangled electron-positron pair created by a strong electric field.
\end{abstract}
\maketitle
\section{Introduction}
\label{RelWaveEqnZB}
The problem of integrating spin and relativistic covariance into a quantum mechanical description of an electron was addressed by Dirac with his relativistic wave equation. Since an electron wave function is described by two independent spin base states, each with positive or negative energy, a solution of Dirac's equation is a four component spinor. The Hamiltonian is  a 4 $\times$ 4 matrix, which for an electron in a  field described by the potentials $(\phi, \textbf{A})$  takes the form \cite{Dirac}
\begin{equation}H = c\boldsymbol{\alpha}\cdot (\textbf{p} - e\textbf{A}) + \beta mc^2 + e\phi \label{DiracHam}\end{equation}
Here $mc^2$ is the electron rest energy, $\textbf{p}$ the electron momentum and $\boldsymbol{\alpha}$ and $\beta$ are 4$\times$4 matrices 
\begin {equation}
\alpha_i = \begin{bmatrix}
0 & \sigma_i \\ 
\sigma_i & 0
\end{bmatrix} \hskip0.3in\beta = \begin{bmatrix}
I & 0\\ 
0 & -I \end{bmatrix}\label{alphabet}\end{equation}
that operate on the $4\times 1$ Dirac spinors. The $\sigma_i$ are the familiar $2\times 2$ Pauli matrices and $I$ is a $2\times2$ identity matrix. It is evident that the $\alpha_i$ and $\beta$ are traceless and hermitian.

While Dirac's equation is fundamental to the highly successful quantum electrodynamics, the interpretation of this equation is an ongoing matter, raising questions such as whether it is a single particle wave equation \cite{BjorkenDrell1, Zee, Penrose, Huang}, and whether the Hamiltonian is local \cite{Mulligan}. The purpose of this article is not to offer a comprehensive interpretation of Dirac's equation. This article will use the Dirac equation to explore the electron's internal structure by focusing on two important consequences of the equation: the anomalous velocity of the electron, and the \textit{zitterbewegung} solutions of the equation.

A velocity operator $\dot{x}_i$ can be defined for the Dirac electron using the relation
$ i\hbar\dot{x}_i = - [H, x_i] $ where $H$ is given by Eq (\ref{DiracHam}). 
It is a straightforward step to show that  \begin{equation} \dot{x}_i = \frac{i}{\hbar} [H, x_i] =c\alpha_i.\label{veloper}\end{equation}
Equivalently, $\dot{x}_i =\frac{\partial H}{\partial p_i} = c\alpha_i $. The eigenvalues of each of the three $\alpha_i$ matrices are $\pm 1,$ and so each component of the velocity operator has the two eigenvalues $\pm c.$  
Dirac provided an explanation for this apparent violation of special relativity:
 
``To measure the velocity we must measure the position at two slightly different times and then divide the change of position by the time interval. (It will not do to measure the momentum and apply a formula, as the ordinary connexion between velocity and momentum is not valid.) In order that our measured velocity may approximate to the instantaneous velocity, the time interval between the two measurements of position must be very short and hence these measurements must be very accurate. The great accuracy with which the position of the electron is known during the time-interval must give rise, according to the principle of uncertainty, to an almost complete indeterminacy in its momentum. This means that almost all values of the momentum are equally probable, so that the momentum is almost certain to be infinite. An infinite value for a component of momentum corresponds to the value of $\pm c$ for the corresponding component of velocity (p 262) \cite {Dirac}.''

In the last sentence Dirac is evidently alluding to the relationship between velocity and momentum of a relativistic electron
\begin{equation} \textbf{v} = \frac{\textbf{p}c}{\sqrt{m^2c^2 + p^2}} \label{velocitymomentum}\end{equation}
whereby every component of velocity becomes $\pm c$ as the corresponding component of momentum goes to $\pm \infty$. But - as Dirac himself acknowledges - the validity of Eq (\ref{velocitymomentum}) for quantum mechanics is not obvious. It has been shown \cite{AhmadWigner} that Eq (\ref{velocitymomentum}) holds in quantum mechanics in agreement with the definition of velocity as $\Delta \textbf{r}/\Delta t$ only when the time interval $\Delta t $ is so long that the initial and final uncertainties of the position are of no consequence,  a requirement that contradicts Dirac's stipulation that the time interval should be very short. A quantum mechanical velocity operator $\textbf{v}$ was defined \cite{OConnellWigner1} according to Eq (\ref{velocitymomentum}) related to the expected values of position measured at an arbitrarily small interval of time $\Delta t$ satisfying 
\begin{equation} \langle \textbf{q} (t)\rangle = \langle \textbf{q} (0)\rangle + \Delta t\langle \textbf{v} \rangle \label{positionvelocity} \end{equation}
where $\textbf{q}(t)$ is a position operator \cite{NewtonWigner, OConnellWigner2}. However, it has been proved \cite{Hegerfeldt} that if one insists on strict localizability (A particle state is said to be localized in V at time $t$ if the probability of finding the particle in V is 1) then causality (If at time $t_0 = 0$ a particle state is localized in V, then there is a constant $r =r_t$, such that, at time $t > 0$, the particle, when translated
by $\vec{a}, |\vec{a}| \geq r_t $,  is not in V) will be violated when applying such a position operator. The problems encountered in attempts to provide an adequate definition of the Dirac velocity operator have led to speculation that any such operator could be no more than a mathematical artifact of Dirac's equation without physical meaning \cite{Bunge}.

It can be readily shown - as we will see in the next section - that the velocity operator has eigenvalues $\pm c$ for an electron state that is a superposition of positive and negative energy states.  In this article we will examine some important consequences of this result. A superposition of positive and negative energy states emerges as a solution to the Dirac equation, and Schr\"odinger labeled the motion of an electron in such a state as zitterbewegung \cite{Schrodinger}. Zitterbewegung has since been recognized as a manifestation of the structure of the electron, specifically as the electron's ``relativistic structure'' \cite{BjorkenDrell1}, though zitterbewegung has always been understood as a feature that is independent of the velocity of the electron, executed even by an electron with zero momentum. An investigation of the internal geometry of the electron undergoing zitterbewegung was carried out by Barut and Bracken \cite{BarutBracken}, who defined a position operator to represent the position of the electron as \textit{a point charge}. We will pursue this line of interpretation. We will identify the velocity operator with the time derivative of this charge position operator and explain why such an identification is justifiable. There is an emerging realization that the center of mass of the electron is not identical with the center of charge \cite{ZouZhangSilenko}, but the relationship between the two has not been established. In this paper we will focus on the center of charge, which we shall call the position of the charge, and show that in zitterbewegung this position of the electron charge is displaced at the speed of light, a result that is both consistent with and required by Special Relativity. We will explore some interesting features of this charge zitterbewegung. We will provide a theoretical formulation, adduce   experimental evidence, and show that our theoretical predictions are borne out in a  model of the creation of an entangled electron-positron pair in a strong electric field.

In the following section we will show (a) that the eigenvalue $c$ of the electron velocity cannot be the velocity of the mass of the electron, and (b) that this velocity describes the motion of the electron charge. 

\section{Theoretical Analysis}
\subsection{Electron Velocity and Electron Mass}
\label{velocitymass}

 It follows directly from (\ref{DiracHam}) and (\ref{alphabet}) that the Dirac Hamiltonian does not commute with any of the velocity operators:
\begin{equation} [H, c\alpha_i] \neq 0 \end{equation}

\noindent Conversely, it is also trivial to verify that an eigenstate of $H$ which is  \begin{equation}\textup{a positive energy state such as}\begin{bmatrix}a
\\ 
b\\ 0\\0\end{bmatrix}  \textup{or a negative energy state such as} \begin{bmatrix} 0
\\ 
0\\ b\\a\end{bmatrix} \nonumber \end{equation} is not an eigenstate of the velocity operator $c{\alpha_x}$ (with corresponding statements for $c\alpha_y$ and $c\alpha_z$).
Moreover, it is easy to show that the eigenfunctions of each velocity operator are superpositions of positive and negative energy states \cite{BjorkenDrell1}. Though it is elementary, we will provide a proof of this result, as this will help clarify the arguments we will make further in this paper.

There are three components of the velocity operator corresponding to the three $\alpha_i$ matrices. A possible set of normalized eigenstates - using the parameters $a$ and $b$ - corresponding to the positive eigenvalue $+c$ of $c\alpha_x$, $c\alpha_y$ and $c\alpha_z$   are, in that order:
\begin{equation} \textup{Eigenstates of} \hskip0.1in c\alpha_x, \hskip0.1in c\alpha_y \hskip0.1in\textup{and}\hskip0.1in c\alpha_z = \begin{bmatrix}a
\\ 
b\\ b\\a\end{bmatrix},\hskip 0.6in  \begin{bmatrix}a
\\ 
b\\ -ib\\ia\end{bmatrix}\hskip 0.3in \textup{and}\hskip0.3in\begin{bmatrix}a
\\ 
b\\ a\\-b\end{bmatrix}\label{eigenstates}\end{equation}
and a similar set can be obtained for the negative eigenvalue $-c.$
It is evident that each of the three eigenstates listed above contains equal proportions of positive and negative energy states, since the probability of finding the electron in  a positive (or  negative) energy state becomes $|a|^2 + |b|^2 = 1/2$. Hence the velocity eigenstates are superpositions of two mutually orthogonal energy eigenstates and therefore not eigenstates of the Hamiltonian of Eq (\ref{DiracHam}). It is common to identify  the ``physically meaningful'' states of the electron with the positive energy eigenstates  $|\psi\rangle$ of the Hamiltonian. These in turn are combinations of the eigenstates of the velocity operators $ |\psi\rangle = a_1|\phi_1\rangle + a_2|\phi_2\rangle $
where $c\alpha_i|\phi_1\rangle = +c|\phi_1\rangle$ and $c\alpha_i|\phi_2\rangle = -c|\phi_2\rangle.$ Thus the expected value of any component of the velocity of an electron has magnitude less than $c$ because $\langle\psi|c\alpha_i|\psi\rangle = (|a_1|^2 - |a_2|^2)c$ which has a magnitude less than $c$ for non-vanishing $a_1$ and $a_2$. The eigenvalues $\pm c$ are thus physically unrealizable asymptotes \cite{Dirac}, and have been labeled as ``internal velocity"\cite{HongSternbergGarcia}, with no physical meaning  ascribed to them, since they do not have definite energy. But we will show that they do have physical meaning, with measurable consequences.  

\subsection{Electron velocity and Zitterbewegung}

We have shown above that if an electron is measured having definite energy (either positive or negative), it cannot be simultaneously observed traveling at the speed of light, and thus Special Relativity is not violated. Only an electron in a state with equal contributions from positive and negative energy states could move at light speed. It was just such a state that was described by Schr\"odinger \cite{Schrodinger} as \textit{zitterbewegung} or ``jittery motion''. The name suggests an electron fluctuating between positive and negative energy while its spatial position also oscillates rapidly. The amplitude of this spatial oscillation was shown to be of the order of the Compton wavelength ($\sim \frac{\hbar}{mc}$) \cite{Dirac, BjorkenDrell1, Mulligan}. Because of the very high frequency of zitterbewegung of the Dirac electron, of the order of $2mc^2/\hbar = 2 \times 10^{21} \sec^{-1},$ a direct observation of this phenomenon has been virtually impossible, and so investigations of zitterbewgung are primarily theoretical in nature \cite{Schrodinger, Dirac, Huang, Mulligan, Sasabe, Welton, Rae, BjorkenDrell1, BarutBracken, RoyBasuRoy, Hestenes, Milonni, Sidharth, KrekoraSuGrobe, WangXiong2, WangXiong, OConnell1,ZouZhangSilenko}.

In contrast to true zitterbewegung, simulations of zitterbewegung have been implemented in the laboratory. The oscillatory motion due to the force exerted on an electron moving in an electric field and the Larmor precession of an electron in a magnetic field are both described as zitterbewegung~\cite{Shen}. Studies on spintronics apply this term to the coupling between the components of the eigenstates of a system~\cite{CsertiDavid}. Trapped ions are also candidates for simulating zitterbewegung~\cite {LamataLSS,GerritsmaKZS,Wunderlich}. Such particles obey Hamiltonians mathematically similar to that of a Dirac electron, such as the Dresselhaus and the Rashba Hamiltonians~\cite{SchliemannLossWestervelt1, SchliemannLossWestervelt2}. But because all these phenomena are qualitatively distinct from the zitterbewegung of the Dirac-Schr\"odinger theory, authors tend to qualify their statements with disclaimers such as ``simulation of the Dirac equation''~\cite{LamataLSS}, ``zitterbewegung-like phenomena''~\cite{ZulickeWB}, ``analogy''~\cite{Zawadski1}, ``reminiscent of zitterbewegung'' or ``zitterbewegung effect''~\cite{JiangLZL}, etc. The essential difference between the simulated zitterbewegung observed in spintronics and the zitterbewegung of the Dirac electron  is that the former is encoded in the spin degree of freedom~\cite{BermudezMDS}, whereas the latter involves an oscillation between positive and negative energy states.  

True zitterbewegung is the rapid oscillation of the electron energy between positive and negative values, accompanied by some sort of rapid spatial oscillation at the speed $c$. But we have shown that this spatial motion cannot be a physical motion of the electron mass. Zitterbewegung is distinct from \textit{quiver motion}, which is the oscillatory motion of a charged object in an oscillating electric field, such as an electron irradiated by a laser~\cite{KrainovSofronov}. Zitterbewegung is also different from  quantum beats of the particle density in propagation and tunneling \cite{FNGJV}. We will next show that zitterbewegung as a translatory motion is to be identified with the motion of the electron charge, quite distinct from any motion of the mass. Thus, this article will prove that charge oscillates at the speed $c$ when the electron is in a zitterbewegung state, without an accompanying mechanical motion of the mass.

\subsection{Electron Velocity and Electron Charge}
\label{velocitycharge}
In the literature of quantum mechanics the conventional particle position of an electron has been identified with the ``mass point'' of the electron \cite{ZouZhangSilenko}. But Barut and Bracken defined a coordinate operator $\textbf{x}$ associated with Dirac's equation explicitly as the ``position of the charge''~\cite{BarutBracken}, which is not necessarily the position of the mass. And they have identified the time derivative of this position operator~\cite{BarutBracken} with the velocity operator obtained by Schr\"odinger~\cite{Schrodinger, Dirac}:
\begin{equation} \frac{d\textbf{x}}{dt} = c\boldsymbol{\alpha}\label{zbwvelocity} \end{equation}
We uphold the validity of this identification by pointing out that, unlike the electron mass, the electron charge does not exhibit any variation with speed. We could therefore identify $-e\textbf{v} \equiv -ec\boldsymbol{\alpha}$ as the charge current operator \cite{BjorkenDrell2}, where $e$ is the magnitude of the negative electron charge. This operator commutes with the velocity operator, thereby indicating that electron charge could be observed traveling at speed $c$. And so we propose that zitterbewegung is the rapid spatial oscillation of electron charge, even as the energy of the electron fluctuates rapidly between positive and negative values. This zitterbewegung of electron charge has measurable consequences.

Upon integrating the velocity operator as a function of time one obtains the time-dependent position operator \cite{Dirac,BarutBracken, Messiah, RoyBasuRoy, Sidharth, ZouZhangSilenko}:
\begin{equation} \textbf{x}(t) = \textbf{a} + c^2 H^{-1} \textbf{p} t + \frac{i\hbar c}{2} (\boldsymbol{\alpha} - cH^{-1} \textbf{p})H^{-1}e^{-2iHt/\hbar}\label{position} \end{equation}
where $\textbf{a}$ is a constant of integration. $c^2H^{-1}\textbf{p}t$ represents the ``classical'' displacement of the electron with velocity $\textbf{v} = \frac{c^2\textbf{p}}{E}$. The remaining``quantum'' term in which $\hbar$ appears as an explicit factor vanishes as $\hbar \rightarrow 0$, and oscillates rapidly with time according to $e^{-2iHt/\hbar}$. This is the zitterbewegung term. So zitterbewegung is the difference between the ``instantaneous velocity'' and the ``average velocity (= $\textbf{p}/\mu$ where $\mu$ is the relativistic mass)'' \cite{Huang}. The part of this term that contains $\boldsymbol{\alpha}$ goes to zero in the non-relativistic limit as $c\rightarrow \infty$. Thus zitterbewegung of the Dirac electron is a relativistic quantum effect, contrary to some claims \cite{Hestenes}. The displacement $\textbf{x}(t) -\textbf{a}$ reduces to this relativistic quantum term $\frac{1}{2} i\hbar c\boldsymbol{\alpha} H^{-1} e^{-2iHt/\hbar}$ in the center of mass sytem of the electron where $\textbf{p} = 0.$ The time dependent position operator $\textbf{x}(t) -\textbf{a}$ in the center of mass system (writing $\textbf{x} (t) - \textbf{a}$ as $\textbf{x} (t)$) becomes  
\begin{equation} \textbf{x}(t) = \frac{1}{2} i\hbar c\boldsymbol{\alpha} H^{-1} e^{-2iHt/\hbar} \label{position}\end{equation}
Since these position operators (one for each component of $\textbf{x}$) are not hermitian, they do not represent observables, an important fact that has appears to have been overlooked by many ~\cite{Dirac,OConnellWigner2,MannMurphy,RoyBasuRoy, Sidharth}. The Dirac particle is thus different from the Schr\"odinger particle, for which the position operator $x_i$ is hermitian and has eigenfunctions in both configuration space and momentum space \cite{BarutMalin}. Because the Foldy-Wouthuysen transform has received some attention in the interpretation of Dirac's equation \cite{BjorkenDrell1,FW, BarutBracken, Messiah, OConnell1,ZouZhangSilenko}, it is instructive to view the Foldy-Wouthuysen position operator:

\begin{equation} \textbf{X} = \textbf{x} + \frac{i\beta\boldsymbol{\alpha}}{2E_p} - \frac{i\beta(\boldsymbol{\alpha}\cdot\textbf{p})\textbf{p} + [\boldsymbol{\sigma}\times \textbf{p}]p}{2E_p(E_p + m)p} \label{fw} \end{equation}
\noindent where \textbf{x} is the non-hermitian Dirac position operator. It is evident that the F-W position operator \textbf{X} is also not hermitian.  

A quest for a position operator yielding a measurable quantity is better served by the hermitian operators $x_i^{\dagger}x_i$   - where $x_i$ is any one component of $\textbf{x}$. Such an operator has been postulated \cite{BarutBracken, BarutMalin}, but its properties have not been adequately explored. In the electron center of mass frame this operator has a particularly simple form. Multiplying one component $x_i$ of Eq (\ref{position}) on the left by its adjoint $x_i^{\dagger}$ we obtain
\begin{equation} x_i^{\dagger} x_i = \frac{\hbar^2c^2}{4H^2}\label{zbwamplitude}\end{equation}
Since $x_i^{\dagger} x_i$ is quadratic in the Hamiltonian it has eigenstates which need not be pure positive (or negative) energy states. They could therefore also be superpositions of positive and negative energy states, such as the velocity eigenstates discussed earlier.  Thus $x_i^{\dagger}x_i$ commutes with both $H$ and $\alpha_i$, and has eigenvalue $\hbar^2c^2/(4E^2)$ where $E$ (or -$E$) is the energy of the electron.

This eigenvalue is in agreement with the well known zitterbewegung amplitude, which for an electron at rest is of the order of its Compton wavelength~\cite{Dirac,Huang, BjorkenDrell1, Mulligan, Sidharth,MannMurphy}.  Eqs (\ref{zbwvelocity}) and (\ref{zbwamplitude}) allow us to provide a physical interpretation of this amplitude. These equations suggest an electron jumping back and forth across a separation of the order of $\hbar/mc$ at the speed $c$ ($\frac{d\textbf{x}}{dt} = c\boldsymbol{\alpha}$), except that in our interpretation it is the charge alone that makes this motion, not the mass. Hence there is no ``physical motion'' associated with zitterbewegung, if by physical motion is meant the motion of the mass \cite{Kurzynski, Dragoman}. And since the continuously varying position operator of Eq (\ref{position}) is not hermitian, it does not represent a continuous displacement of the charge, and hence there is no charge acceleration. The hermitian operator $x_i^{\dagger}x_i$ measures the square of a distance, and hence the magnitude of the distance. The electron charge apparently covers this distance at the speed of light. So, according to this model, the charge disappears at one point in space-time and reappears at another point which has a time ordered light-like separation from the first. The electron charge displaced at light speed from one point to another generates and experiences a  potential energy as the charge interacts with itself via a Li\'enard-Wiechert potential \cite{Keller, Jackson}. The potential at a point $\textbf{r}$ at time $t$ due to a charge $Q$ at a point $\textbf{r}_{Q(t')}$ at time $t'$ is given by 
\begin{equation} \phi(\textbf{r}, t) = \frac{Q}{4\pi \epsilon_0}\int_{-\infty}^{\infty} \frac{1}{|\textbf{r} - \textbf{r}_Q(t')|} \delta\left(\frac{|\textbf{r} - \textbf{r}_Q(t')|}{c} - t + t'\right) dt' \end{equation}
Since the zitterbewegung electron is separated from its ``twin'' by a light-like interval, we could picture ``one'' electron at $\textbf{r}$ at time $t$ experiencing the potential of the ``other'' electron at $\textbf{r}_Q$ at time $t'$ such that $|\textbf{r} - \textbf{r}_Q(t')| = c (t - t')= \frac{\hbar}{mc}$. So in the electron center of mass system the length $|\textbf{r} - \textbf{r}_Q(t')|$ is the eigenvalue of the operator $\sqrt{x_i^{\dagger}x_i}$. The delta function reduces the integral to 
\begin{equation} \phi (\textbf{r}, t)  = -\frac{mce}{4\pi\epsilon_0 \hbar} \end{equation}
 \noindent Thus zitterbewegung has the effect of increasing the electron energy by an amount equal to 
$$-e\phi(\textbf{r}, t) = \frac{mce^2}{4\pi\epsilon_0 \hbar} = \frac{mc^2e^2}{4\pi\epsilon_0 \hbar c} = mc^2\alpha$$
The total energy of the electron is therefore $mc^2(1 + \alpha)$ and the electromagnetic mass becomes $m(1 + \alpha)$. So there is an explicit causal relationship between zitterbewegung and the electromagnetic mass of the electron. This is in agreement with the relationship obtained by Schwinger of the ``mechanical mass'' $m$ to the ``electromagnetic mass $m (1 + \alpha)$'' of the electron \cite{Schwinger1} due to interaction with the field.

We would obtain the same result if we model the electron as a uniformly charged spherical shell of diameter $\hbar/mc$ (with fields inside the shell canceled by the zitterbewegung positions of the electron), and then calculate the total energy of the electromagnetic field due to this spherical shell. This latter energy comes to $\alpha mc^2$. This model is to be contrasted with Casimir's semi-classical model of the electron as a solid spherical shell whose radius can be made asymptotically equal to zero \cite{Casimir, PuthoffC} by assuming a negative interior Casimir energy that balances the external positive electromagnetic energy. Casimir energy need not be invoked in our model since there is no solid shell involved.  

It has been argued that since the Foldy-Wouthuysen transform eliminates zitterbewegung, the latter cannot be an observable \cite{OConnell1,ZouZhangSilenko}. We have shown that this would be true only if zitterbewegung is understood as a motion of electron mass, which it is not.

\subsection{Zitterbewgung and Field Theory}
\label{qft}
 J. J. Sakurai conjectured that zitterbewegung arises from the influence of virtual electron-positron pairs (or vacuum fluctuations) on the electron \cite{Sakurai}. W. Thirring proposed that in zitterbewegung the original electron annihilates the positron of the virtual pair, leaving the other electron at a distance relative to the original electron \cite{Thirring}. In a more recent theoretical study, which we shall discuss presently in more detail, Krekora, Su and Grobe have shown that zitterbewegung cannot occur apart from the creation and annihilation of virtual electron-positron pairs in the vacuum field~\cite{KrekoraSuGrobe}, a result corroborated by Wang and Xiong~\cite{WangXiong}. According to this field theoretical model, the physical electron annihilates the virtual positron, thereby making the virtual electron real. Krekora \textit{et al} have shown that the electron and the positron are not created at the identical spatial point, but that in a weak or zero electric field there is a distance of the order of the Compton wavelength between these particles. (The distance decreases with the strength of the field.) Milonni has described the process in terms of electrons interacting with holes in the negative electron sea \cite{Milonni}. Either way, when the electron annihilates the virtual positron (or fills the hole) the new electron appears at a distance from the original electron, the distance being of the order of the Compton wavelength and reducing to zero as the field strength goes to infinity.  Thus, a field theoretical model of zitterbewegung can be described as virtual electron-positron pairs continuously created and annihilated around a real electron, and the original electron annihilates with the virtual positron, leaving the (newly created) electron as the real particle.  

The insight that we contribute to these stated results is the emphasis that it is the charge that is transported spatially, not the mass. The charge of the real electron neutralizes the charge of the virtual positron, and reappears at a spatially displaced position as the charge of the former virtual electron.  A repetition of the process could bring the electron charge back to the original position, or to a different position, the directions being determined randomly. 

We showed above in Sec \ref{velocitycharge} that the zitterbewegung displacement at the speed of light is a relativistic quantum effect. We then identified this displacement as a displacement of the electron charge. We shall now show that this charge zitterbewegung is consistent with a relativistic quantum field theory.
 
A free electron - represented by a plane wave - that is initially in a state of positive (negative) energy, remains in a positive (negative) energy state, but in the presence of an electromagnetic field the wave function becomes a wave packet containing both positive and negative energy states~\cite{BjorkenDrell1}. But since even the vacuum contains zero point energy, a free electron is never in purely vacant space~\cite{BjorkenDrell1, Milonni}. The electron interacts with the vacuum and undergoes zitterbewegung. The spontaneous appearance and disappearance of virtual electron-positron pairs in the vacuum requires that $\nabla\cdot \textbf{E} \neq 0$ within a small spatial volume (of dimensions less than of the order of the Compton wavelength) at any given time even if $\langle \Phi|\nabla\cdot\textbf{E}|\Phi\rangle = 0$ everywhere \cite{KrekoraSuGrobe, WangXiongQiu}. (Here the divergence of the vacuum field charge has been eliminated by a suitable normal ordering \cite{BjorkenDrell2}.) In Maxwellian electrodynamics an electric field propagates at speed $c$ in the absence of charges:

\begin{equation} \nabla^2 \textbf{E} = \frac{1}{c^2}\frac{\partial^2 \textbf{E}}{\partial t^2} \label{waveE} \end{equation}
In the standard derivation of this equation it is assumed that $\nabla\cdot \textbf{E} = 0$ everywhere in the space under consideration. But in a quantum description this requirement is replaced by $\langle \Phi|\nabla \cdot \textbf{E}|\Phi\rangle = 0$ without demanding that $\nabla\cdot \textbf{E} = 0$. So, permitting  $\nabla \cdot \textbf{E} = \rho/\epsilon_0$  (for the electron charge undergoing zitterbewegung) and taking the divergence of both sides of Eq (\ref{waveE}) we obtain $\nabla^2 \rho = \frac{1}{c^2}\frac{\partial^2\rho}{\partial t^2}$. This equation shows that in zitterbewegung electric charge is transported at light speed.  

 The Aharanov-Bohm effect implies that charges interact  directly not with fields but with potentials~\cite{AharonovBohm, TOMKEYY, OMKET, PeshkinTonomura, Tonomura,VieiraBakke}. So the Aharonov-Bohm effect has moved ``the scalar and vector fields from the category of mathematical fictions to the category of physical ontology''~\cite{Maudlin}, a claim which finds some justification in the gauge-invariance of the Aharonov-Bohm phase~\cite{Cohenetal}. The force acting on a charge in a potential $\phi$ created by static charges equals -$q\nabla \phi$, and in order to experience this force the charge will have to ``measure'' not the scalar potential but the gradient of the potential at the point that it occupies. And a measurement carried out entirely at one point in space cannot yield a spatial derivative of the potential at that point, since by definition $\frac{d\phi}{dx} = \lim_{\Delta x\rightarrow 0} \frac{\Delta \phi}{\Delta x}$ and the derivative does not exist unless the function $\phi$ can be measured at two distinct points at an infinitesimal displacement $\Delta x$ relative to each other. So, in classical physics, a point charge - like a point mass in a gravitational field - must be considered as the limiting case of a finite object. Thus, the notion of a pure point charge is an impossible idealization in classical physics. It is a mathematical convenience to consider charge densities of the form $q\delta (x - a)$~\cite{Jackson}, but it is evident that a stationary point charge will not experience a force in an electric field, unless the point is understood as the limiting case of a shrinking volume. But the point charge of an electron is not generated by shrinking a classical charged sphere to a mathematical point, for such a shrinking would entail an infinite increase of electromagnetic self-energy. So, an electron charge is just a point charge with zero dimension. And in order to measure the electric field in any direction a point charge must evaluate the electric potential at two different points in space \textit{simultaneously}. Simultaneity implies a light-like separation, which requires a teleportation of charge at the speed of light between the points. Space-like separations are forbidden by Special Relativity. A time-like separation will not work, for one could always transform via a Lorentz boost to a reference frame where the two points have different time coordinates but the same spatial coordinate, and so the gradient of the potential would not exist at that spatial point in that frame \cite{Jackson}. Hence it is necessary that a point charge should ``scan'' the space of the electric potential at the speed of light, which remains invariant under Lorentz boosts.

Zitterbewegung permits a stationary electron - i.e. one with zero center of mass momentum - to scan the scalar potential in any direction. Experimental evidence for this, including the magnitude of the zitterbewegung displacement of $\frac{\hbar}{mc}$, is to be found in the Darwin term correction to the Coulomb potential experienced by an electron in an atom \cite{BjorkenDrell1}:
\begin{equation} \langle \delta V\rangle = \langle V (\textbf{r} + \delta \textbf{r})\rangle - \langle V(\textbf{r})\rangle = \left\langle \delta r \frac{\partial V}{\partial r} + \frac{1}{2} \sum_{ij} \delta r_i\delta r_j \frac{\partial^2 V}{\partial r_i\partial r_j}\right\rangle \cong \frac{1}{6} \delta r^2 \nabla^2V \simeq \frac{\hbar^2}{6m^2 c^2}\nabla^2 V\nonumber \end{equation}
	In the above equation it is the motion of the charge and not of the mass that enters the calculation, since ``electromagnetic interactions are defined by the center-of-charge position but not by the center-of-mass'' position \cite{ZouZhangSilenko}.

\section{Zitterbewegung of Entangled Electron-Positron Pair}
\label{entangle}
Electron-positron pair production by the application of a strong electric field to the vacuum is theoretically possible, even if direct experimental verification of such a process is not yet available \cite{LvLiSuGrobe}. Krekora, Su and Grobe \cite{KrekoraSuGrobe} have carried out an analysis of the creation of an electron-positron pair in the presence of a potential that is turned on at time $t = 0$. The authors present their results as evidence of the absence of zitterbewegung understood as a bodily motion of electron mass. But we have shown that zitterbewegung is not the motion of mass, but successive displacements of charge.  And so we will show that the results obtained by the authors - which we shall henceforth refer to as the Study - actually confirm our predictions regarding the motion of electron charge in zitterbewegung.

The authors have analyzed the ``birth process'' of an entangled electron-positron pair in the presence of a potential of the form 
\begin{equation} V (x,t) = V\left[ \frac{\tanh(x/W) + 1}{2}\right] \label{potential}\end{equation} that is turned on at time $t$ = 0. The absence of an expected $\theta (t)$ factor suggests that their interest is not in the actual process of the switching on of the potential. They have set up equations for the entangled electron-positron wave function for which they obtained analytic and numerical solutions which they have displayed graphically. The entangled electron-positron wave function $\phi (x, y, t)$ is obtained by solving the Dirac operator equation (written in a. u.):
\begin{equation} i\partial_t \hat{\Psi} (x, t) = [c\boldsymbol{\alpha}\cdot\textbf{p} + \beta c^2 + V(x,t)]\hat{\Psi} (x, t)\end{equation}

\noindent and then setting up the entangled wave function as
\begin{equation} \phi(x, y, t) = \langle 0|| \hat{\Psi}^{(+)}(x,t) \hat{\Psi}^{(+)}_c (y,t)||0\rangle \end{equation}
where the (+) superscript indicates that only positive frequencies are employed in the field operators. The electron position probability density is then defined as
\begin{equation}\rho_0 (x) = \int dy \rho_0 (x,y)\end{equation} where $x$ is the position of the electron and $y$ the position of the positron. The unitary time development of the operators is performed by applying the operator

\begin{equation}U(t) \equiv \exp \left( -i[c\boldsymbol{\alpha}\cdot\textbf{p} + \beta c^2 + V(x)]t\right)\end{equation}
to each possible initial state. It is evident that the potential has the effect of linking positive and negative energy states and thereby placing the electron (and the positron) in a state that includes zitterbewegung \cite{BjorkenDrell1, WangXiong}. Contour plots of the electron-positron density $\rho (x, y, t)$ obtained by Krekora et al in their Study are displayed in Fig 1.

\begin{figure}
	\centering
		\includegraphics[width=6in]{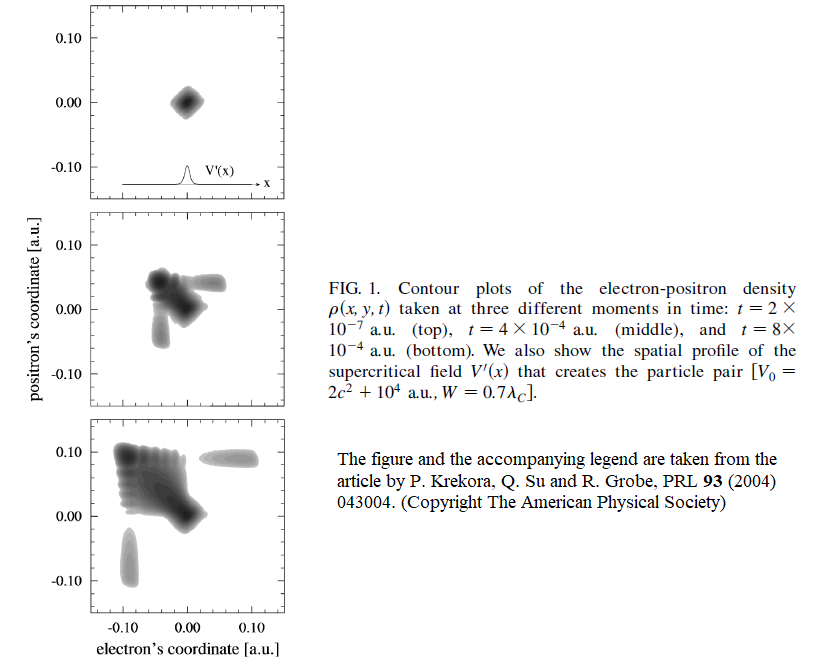}
		 
\label{fig A} 
\end{figure}

1. The electron-positron wave functions obtained in the Study contain both positive and negative energy states.  The non-vanishing wave functions include positive energy ($\phi_0^{1,2}$), negative energy ($\phi_0^{3,4}$), as well as coupling between particles with opposite energy ($\phi_0^{1,4}$ and $\phi_0^{2,3}$), and the states obtained by switching the superscripts (electron and positron). So these states include entangled electron-positron positive-negative energy states: $\phi_0^{1,4}, \phi_0^{4,1}, \phi_0^{2,3}, \phi_0^{3,2}$.  The electron-positron entangled state is therefore a mixture of regular states of positive energy and negative energy, and entanglements of positive and negative energy states. The presence of these states indicates zitterbewegung of the electron-positron pair. This is consistent with the well-known result that any electron (or positron) in an electromagnetic field is in a superposition of states including zitterbewegung states \cite{BjorkenDrell1}.

2. The electron density $\rho_0 (x)$ depicted in  Fig 3 of the Study shows gaussian shapes for different field strengths. A gaussian position wave function $\psi (x)$ implies that $\langle p_x\rangle = 0.$ So there is no bodily motion of either the electron or the positron after their creation.  Thus, according to the earlier discussion of Eq (\ref{position}) the electron density $\rho_0 (x) = \int dy \rho_0 (x,y)$ yields the charge density, not the mass density. Our prediction was that the electron charge jumps discretely between two points without flowing continuously between them. An examination of the lowest diagram in Figure 1 shows a clear spatial gap between the main body and the two ``satellites,'' confirming our prediction that the electron charge density jumps discontinuously between the two ``satellites,'' which are here centered at $\pm (-0.04, 0.04) $ a.u. 

3. In Section \ref{velocitycharge} we showed that the amplitude of zitterbewegung is of the order of the Compton wavelength. In the Study the amplitude - which is the separation between the main body and each satellite - is of the order of 0.05 a.u. = $2.6 \times 10^{-12}$ m, which is of the order of the Compton wavelength ($2.4\times 10^{-12}$ m). 

4. We showed that in zitterbewegung the electron charge leaps from one extreme point to the other at the speed of light. We will now calculate the speed with which the particles leap between the ``satellites'' in the Study. Since the time is $8\times 10^{-4}$ a.u. = $1.9 \times 10^{-20}$ s, and the spatial separation is of the order of 0.10 a.u. = $5.3\times 10^{-12}$ m, the speed of the electron comes to $2.8\times 10^8$ m/s, which is the speed of light within the limits of accuracy. 

5. The Study showed that in the narrow field ($W \rightarrow 0$) the electron's spatial uncertainty cloud becomes arbitrarily narrow, and hence there is no lower limit to the electron's localization. For the case of the supercritical field produced in the limit $W\rightarrow 0$ in the potential $V(x,t) = V[\tanh (x/W) + 1]/2$, the operator $H^2 \rightarrow \infty$ and so our position operator $x^{\dagger}x \rightarrow 0$ according to Eq (\ref{zbwamplitude}). Thus, as the field strength goes to infinity, the amplitude of the electron's zitterbewegung motion goes to zero. So, according to the field theoretical model outlined in Sec \ref{qft}, the distance between the created electron and positron also goes to zero in the strong field limit. 

 \section{Velocity of the mass of the electron}

In this paper we have focused almost exclusively on the position and velocity of the electron charge. In Eqn (\ref{position}) the position of the mass was denoted by the term $c^2H^{-1}\textbf{p} t$ which we identified with the ``classical'' position of the center of mass $\textbf{v}t$. But there is another position term also containing momentum: $-\frac{i\hbar c^2}{2} H^{-1}\textbf{p} H^{-1}e^{-2iHt/\hbar}$. This is an oscillatory term of frequency $2H/\hbar$. This appears to be a zitterbewegung motion of the electron mass, in the direction of the momentum, which could be called ``longitudinal zitterbewegung'' \cite{WangXiong}. But this is qualitatively different from the zitterbewegung of the electron charge, for the following reasons.

A study of the motion of the mass would have to include gravity as well, since by the Principle of Equivalence the inertial mass of an object is equal to its gravitational mass. The motion of an object having charge $q$ and mass $m$ in an electromagnetic and a gravitational  field can be described by an equation such as \cite{Carroll}
\begin{equation} m\frac{d^2 x^{\sigma}}{d\tau^2} = q F^{\sigma}_{\nu} \frac{dx^{\nu}}{d\tau} - m\Gamma^{\sigma}_{\mu \nu} \frac{dx^{\mu}}{d\tau}\frac{dx^{\nu}}{d\tau}  \label{gravity}\end{equation}
where $\tau$ is the proper time, $F^{\sigma}_{\nu}$ the electromagnetic field tensor and  the Christoffel symbol $\Gamma^{\sigma}_{\mu \nu} = \frac{1}{2} g^{\sigma \rho} [ \partial_{\mu} g_{\nu \rho} + \partial_{\nu} g_{\rho \mu} - \partial_{\rho} g_{\mu \nu}]$ expresses the curvature of spacetime due to gravity. In Sec \ref{qft} we showed that the spatial derivative of the electric potential requires that a point charge should scan the electric field through zitterbewegung. In the second term on the right side of Eq (\ref{gravity}) the mass is multiplied by derivatives of the metric (through the $\Gamma^{\sigma}_{\mu \nu}$ term). So, by analogy, a point mass must scan the gravitational field via a sort of zitterbewegung in order to experience a gravitational force. But here is a problem. The term $-\frac{i\hbar c^2}{2} H^{-1}\textbf{p} H^{-1}e^{-2iHt/\hbar}$ does not represent motion at the speed of light, and our discussion in Sec \ref{qft} showed that the scanning should take place at the speed of light. Wang, Xiong and Qiu \cite{WangXiongQiu} have developed a theory of photonic zitterbewegung and have put forth the hypothesis that whereas electron zitterbewegung is connected to the electromagnetic field, photonic zitterbewegung is connected to the gravitational field. A possible corollary to this hypothesis would be that photons play an intermediary role in the scanning of a gravitational field by an electron. We will explore this line of investigation in a subsequent publication.

 \section{Conclusions}

We have shown that the velocity operator $c\boldsymbol{\alpha}$ can be identified with the velocity of the electron charge or center of charge, and that any component of this operator has eigenvalues $\pm c$. We identified this as the velocity with which the electron charge is displaced through a spatial distance of the order of the Compton wavelength or less for higher energies. Our field theoretical analysis showed that zitterbewegung can be described as a real electron charge interacting with a virtual electron-positron pair of charges, annihilating the virtual positive charge and continuing its existence as the virtual electron charge become real. 

Our model of zitterbewegung requires a revision of theories that base themselves on earlier interpretations of zitterbewegung as a bodily motion of electron mass, cf. \cite{Huang}. Puthoff~\cite{Puthoff} has developed the theory put forth by Sakharov~\cite{Sakharov}, who suggested that the rest mass energy of a particle (called a parton) arises from the mechanical energy of its zitterbewegung motion. But our analysis of the Dirac Hamiltonian has shown that zitterbewegung cannot be the motion of the mass of the electron, and hence there is no kinetic energy associated with zitterbewegung. Thus any physical theory based on mechanical energy due to massive motion in zitterbewegung requires radical revision.

\end{document}